\documentstyle[12pt,aaspp4]{article}

\def\T1{\ {$T_1$}\ }
\def\MT1{\ {$M_{T_1}$}\ }
\def\ct1{\ {$(C-T_1)$}\ }
\def\CT10{\ {$(C-T_1)_0$}\ }
\def\MI{\ {$M_I$}\ }
\def\vi{\ {$(V-I)$}\ }
\def\VI0{\ {$(V-I)_0$}\ }
\def\feh{\ {[Fe/H]}\ }
\def\Wash{\ {Washington}\ }
\def\met{\ {metallicity}\ }
\def\mets{\ {metallicities}\ }
\def\cl{\ {cluster}\ }
\def\cls{\ {clusters}\ }
\def\mp{\ {metal-poor}\ }
\def\mr{\ {metal-rich}\ }
\def\red{\ {reddening}\ }
\def\reds{\ {reddenings}\ }
\def\cal{\ {calibration}\ }
\def\disp{\ {dispersion}\ }
\def\2cd{\ {two-color diagram}\ }
\def\st{\ {standard}\ }

\def\sgb{\ {standard giant branch}\ }
\def\sgbs{\ {standard giant branches}\ }
\def\tech{\ {technique}\ }
\def\dm{\ {$(m-M)_V$}\ }
\def\phot{\ {photometry}\ }
\def\photic{\ {photometric}\ }
\def\pe{\ {photoelectric}\ }
\def\pop{\ {population}\ }
\def\pops{\ {populations}\ }
\def\mag{\ {magnitude}\ }
\def\mags{\ {magnitudes}\ }
\def\sens{\ {sensitivity}\ }

\def\ell{\ {elliptical}\ }

\def\gtsim{\ {\raise-0.5ex\hbox{$\buildrel>\over\sim$}}\ }
\def\ltsim{\ {\raise-0.5ex\hbox{$\buildrel<\over\sim$}}\ }

\begin{document}

\title{Standard Giant Branches in the Washington Photometric System}

\author{Doug Geisler\altaffilmark{1}}
\affil{Kitt Peak National Observatory, National Optical Astronomy 
Observatories,\\
P.O. Box 26732, Tucson, Arizona 85726 \\
{\it dgeisler@noao.edu}}

\author{and}

\author{Ata Sarajedini\altaffilmark{2}}
\affil{San Francisco State University, Dept. of Physics and Astronomy,\\
1600 Holloway Ave., San Francisco, CA 94132 \\
{\it ata@stars.sfsu.edu}}

\slugcomment{Submitted to the Astronomical Journal}

\altaffiltext{1}{Visiting Astronomer, Canada-France-Hawaii Telescope, 
operated by the National Research Council of Canada, le Centre National 
de la Recherche Scientifique de France, and the University of Hawaii.}
\altaffiltext{2}{Hubble Fellow}

\begin{abstract}

We have obtained CCD \phot in the \Wash system $C,T_1$ filters for some 850,000
objects associated with 10 Galactic globular clusters and 2 old
open clusters. These
clusters have well-known metal abundances,
spanning a \met range of 2.5 dex from \feh$\sim -2.25$ to $+0.25$ at
a spacing of $\sim 0.2$ dex. Two independent observations were obtained for
each cluster and internal checks, as well as external comparisons with existing
photoelectric \phot, indicate that the final colors and magnitudes have overall
uncertainties of \ltsim 0.03 mag. 

Analogous to the method employed by Da Costa and Armandroff (1990, AJ, 100, 
162) for $V,I$ \phot, we then proceed to construct standard (\MT1, \CT10) giant
branches for these \cls adopting the Lee et al. (1990, ApJ, 350, 155) distance
scale, using some 350 stars per globular \cl to define the giant branch.
We then determine the \met \sens of the \CT10 color at a given \MT1
value. The \Wash system technique is found to have three times the \met \sens
of the $V,I$ technique. At \MT1$=-2$ (about a \mag below the tip of the giant
branch, roughly equivalent to $M_I=-3$), the giant branches of 47 Tuc and M15
are separated by 1.16 \mags in \CT10 and only 0.38 \mags in \VI0. Thus, for a 
given photometric accuracy, \mets can be determined three times more precisely
with the \Wash technique. We find a linear relationship between  
\CT10 (at \MT1$=-2$) and \met
(on the Zinn 1985, ApJ, 293, 424 scale) exists over the full \met range,
with an rms of only 0.04 dex.
We also derive \met calibrations for \MT1$=-2.5$ and $-1.5$, as well as for
two other \met scales. The \Wash technique retains almost the same \met \sens
at faint \mags, and indeed the standard giant branches are still well
separated even below the horizontal branch.
The \phot is used to set upper limits in the range 0.03 -- 0.09 dex for any
intrinsic \met dispersion in the calibrating \cls.
The calibrations are applicable to objects with ages \gtsim 5 Gyr -- any 
age effects are small or negligible for such objects.

This new technique is found to have many advantages over the old two-color
diagram technique for deriving \mets from \Wash \phot. In addition to only 
requiring 2 filters instead of 3 or 4, the new technique is generally much
less sensitive to \red and photometric errors, and the \met \sens is many times 
higher. The new technique is especially advantageous for \mp objects. The five
\mp clusters determined by Geisler et al. (1992, AJ, 104, 627),
using the old technique, to be much more
\mp than previous indications, yield \mets using the new technique which are in
excellent agreement with the Zinn scale. The anomalously low \mets 
derived previously are
undoubtedly a result of the reduced \met \sens of the old technique at low 
abundance. However, the old technique is still competitive for \mr objects
(\feh \gtsim --1).

We have extended the method developed by Sarajedini (1994, AJ, 107, 618)
to derive simultaneous \red and \met determinations from the shape of the 
red giant branch, the \T1 \mag of the horizontal branch, and the apparent \ct1
color of the red giant branch at the level of the horizontal branch. This
technique allows us to measure \red to 0.025 \mags in $E(B-V)$
and \met to 0.15 dex. Reddenings can also
be derived from the blue edge of the instability strip, with a
similar error.

We measure the apparent \T1 \mag of the
red giant branch bump in each of the calibrating
clusters and find that the difference in \mag between the bump and the 
horizontal branch is tightly  and sensitively
correlated with \met,
with an rms dispersion of 0.1 dex. This feature can therefore also be used to
derive \met in  suitable objects. Metallicity can be determined as well
from the slope of the RGB, to a similar accuracy. Our very populous
color-\mag diagrams reveal the 
asymptotic giant branch bump in several clusters.

Although \MT1 of the red giant branch tip
is not as constant with \met and age as \MI, it is still found
to be a useful distance indicator for objects with \feh \ltsim --1.2. For the
6 standard \cls in this regime, $<M_{T_1}(TRGB)>=-3.22\pm0.11(\sigma)$, with 
only a small \met dependence. This result is found to be in very good agreement
with the predictions of the Bertelli et al. (1994, A\&AS, 106, 275) isochrones.
We also note that the \Wash system holds great potential
for deriving accurate ages as well as \mets.
 
\end{abstract}

\keywords{Galaxy: abundances --- globular clusters: general}

\section{INTRODUCTION}

The color of the first ascent, red giant branch (RGB) of an old stellar system
has long been recognized as a sensitive indicator of metal abundance
(e.g. Sandage and Smith 1966,
Hartwick 1968, Rood 1978, Frogel et al. 1983).
Observationally, the 
utility of this feature for determining \met was first exploited          
using the traditional
Johnson BV filters by Hartwick (1968) in his definition of the $(B-V)_
{0,g}$ feature: the \red-corrected color of the RGB at the level of the 
horizontal branch (HB). Searle and Zinn (1978) formed an abundance ranking
from plots of $M_V$ vs. a reddening-independent color derived from spectral 
scans of globular cluster (GC) giants.

Da Costa and Armandroff (1990,  hereafter DCA) extended
this technique to the $V,I$(Cousins) \photic
system, quoting a large color baseline
as one of their motivators. They utilized the entire upper RGB,
establishing standard GC giant branches in the (\MI,\VI0) plane.
They demonstrated the substantial \met \sens of the \VI0 color at a given \MI.
This method now enjoys great   popularity
as the preferred technique for, e.g.
deriving \mets from \photic   observations of the stellar \pops in distant GCs
and nearby resolved galaxies. Sarajedini (1994) further demonstrated that these
\st RGBs could be used to determine both \red and \met simultaneously, making
the technique of even greater utility. Independently, a number of 
investigators (e.g. Lee et al. 1993)
have shown that the absolute I \mag of the tip of the RGB (TRGB)
is quite insensitive to \met and age over a wide range of both of these
parameters, and can therefore be used as an accurate distance indicator, 
further enhancing the utility of $V,I$ \phot.

Given the distinct advantages of such a technique in many applications, the 
development of a similar technique using other filter combinations, including
different photometric
systems, is warranted. Geisler (1994) and Geisler and 
Sarajedini (1996) first introduced an analogous \tech based on the \ct1 color
of the \Wash \photic system. The \Wash system (Canterna 1976a) is a four filter
broadband system
designed (Wallerstein and Helfer 1966) to provide an efficient yet accurate
measurement of abundances and temperatures for G and K giants. In this regard,
the system has proven useful for determining \mets of individual giants in a 
variety of applications (cg. Geisler et al. 1991,  hereafter GCM). 
GCM showed that the traditional two-color diagram \tech
utilizing \Wash filters offered a powerful combination of efficiency and 
accuracy for determining \met in late-type giants over the full range of
stellar abundances, though with decreased sensitivity for very cool, metal-poor 
stars, as is typical 
of similar techniques.
The system has also proven useful 
for deriving the \met of distant GCs from their integrated color (cg. Harris
and Canterna 1977, Geisler et al. 1996). In this latter application, the \ct1
color has been successful because of its very high \met \sens compared
to other indices such as $(B-V)$ and \vi (Geisler et al. 1996). The \ct1
index enjoys an even wider color
baseline than \vi while still falling within normal CCD response curves.
Both filters are broad, with FWHM $>1000\AA$.
The C filter is similar to the Johnson U filter, including the many spectral
features in the blue-uv from $\sim 3500 - 4500\AA$ that are \met sensitive in cool
giants, but it is significantly broader and redder, making accurate \phot much
more tractable, and \red and atmospheric
extinction less problematic. The \T1 filter is virtually 
identical to $R_{KC}$ (indeed,   
Geisler 1996 has recommended the substitution of the R(Kron-Cousins) filter
for the \T1 filter, in particular because of its much higher efficiency)
and offers a mostly continuum filter near the peak flux
in cool giants and allows a wide color baseline in combination with $C$.
Although the \Wash system includes two other filters, $M$ and $T_2$, the
desirability of deriving a \tech, such as that of DCA, that uses only two
filters to maximize telescope efficiency, is clear. 

In view of the advantages outlined above,  especially the efficiency and
high \met \sens, and the fact that the \ct1
index was already in use for the study of integrated GC colors, we have 
embarked on this investigation to establish \st giant branches in the \Wash
system using the \ct1 color and \T1 \mag.
This work attempts to follow
the high standards of photometric quality and analytical technique
exemplified  by the work of DCA. Throughout this paper, we compare our work to 
that of DCA because the latter has developed into a benchmark dataset
used by many other investigators.
Initial efforts were reported by Geisler (1994)
and Geisler and Sarajedini (1996). Both of these papers demonstrated that such
a \tech has great potential for determining metal abundances in distant objects,
with a \met \sens far
exceeding that of \vi. However, these investigations were
preliminary, with only a limited number of \st \cls, and photoelectric
\phot for only a small
number of stars per \cl.

This paper reports on the establishment of \sgbs in the \Wash system, based on
CCD \phot for thousands of stars in each of
a large number of well-studied Galactic GCs and old open \cls. The 
extensive \phot is 
described in Sect. 2. In Sect. 3, we present the \sgbs. The \met 
calibration is detailed in Sect. 4, where we also compare the \Wash \sgb \tech
with the existing \Wash \2cd \tech and the \vi \sgb \tech. 
In Sect. 5, we develop the \tech to
simultaneously determine \red and \met. Sect. 6 describes how \met can also
be determined from the RGB bump and slope, \red from the 
color of the red edge of the blue HB
and distance from the TRGB.
We summarize our results in Sect. 7.

\section{PHOTOMETRY}

\subsection{Sample Selection}

The choice of which \cls to select as `standard' \cls
is of paramount importance.
Our primary sample selection criteria included: covering the full \met range of
\cls in the Galaxy, with one \cl every $\sim 0.2$ dex in \met, using 
well-studied \cls with accurate \mets, (small)
distances and (low) \reds. The \st \cls
are listed in Table 1, which gives the NGC number, \met and \red (from Zinn
1985 - hereafter Z85 - for the GCs except for NGC1851 whose \met has been 
revised, - see Sect. 4), $(m-M)_V$ and \feh on two other \met
scales. The distances will be 
discussed in Sect. 
3. Note that this sample generally fulfills our criteria quite
well. There are 10 GCs, ranging in \met (on the Z85 scale) from --2.15 to --0.3.
The \reds are generally low, except for NGC5927. Note that 5 of these \cls --
NGC7078 (M15), NGC6397, NGC6752, NGC1851 and NGC104 (47 Tuc) -- are the same
as used by DCA.

In order to cover even higher \mets, we elected to use several 
well-studied old open clusters, NGC2682 (M67) and NGC6791. Although 
approximately solar \met GCs exist, they are very highly reddened and
their distances and \mets are only poorly determined. In contrast, these
parameters are well known for our two
selected open clusters. A potential problem in using open \cls is that age
effects may become important if their ages are significantly less than those
of the GCs. In this respect, NGC6791 is a perfect choice as it is one of the
oldest open \cls known, with an age of $\sim 10$ Gyr (e.g. Tripicco et al. 1995)
and thus quite comparable to those of the GCs in our sample. It also has a
very high \met and is quite populous, allowing for accurate definition of the 
RGB. The case of NGC2682 is less fortuitous: it is only $\sim 4$ Gyrs old (e.g.
Dinescu et al. 1995). Thus, age effects may start to play a role. However, it
represents the oldest, nearest, most populated  and best studied \cl of about
solar \met. Possible age effects will be discussed in Sect. 4.
The \reds, \mets 
and distances we have 
used for NGC2682 and NGC6791 represent means from a number of
studies. Note that Twarog et al. (1997), in their recent compilation and 
homogenization of open \cl properties, give $E(B-V)=0.04$, \feh$=0.00\pm0.09$
and $(m-M)_V=9.68$ for NGC2682 and 0.15, $0.15\pm0.16$ and 12.94 for the 
respective parameters of NGC6791. All of these values are very similar to ours
except for the last one.

We then have 12 \st \cls, double the number used by DCA. Also, our
\st \cls extend to \mets 1 dex more \mr than those of DCA. At the same time,
there is substantial overlap among the \st \cls of both samples, allowing for
an accurate comparison of their respective advantages and disadvantages.

\subsection{Observations}

The observations were secured on a total of 14 nights from April, 1989 to
December, 1996 using the CTIO 0.9m (7 nights), KPNO 4m (3 nights), KPNO 0.9m
(3 nights) and CFHT (1 night). A variety of CCDs was used but generally a 
Tektronix $2048\times2048$ was the detector. The scale was typically $\sim 0.5
\arcsec$/pixel. Several different prescriptions for the \Wash C and
\T1 filters were
used over this extended time period. The recommended prescription 
for C is that given
in Geisler (1996): 3mm BG3 $+$ 2mm BG40. For \T1, we used existing \T1 filters
as well as a \st $R_{KC}$ filter.
Geisler (1996) has shown that this filter is a more 
efficient substitute for \T1 and that $(C-R)$ accurately reproduces \ct1 over
the range $-0.2$\ltsim\ct1\ltsim3.3. A few of the brightest, coolest giants
in the most \mr and/or highly reddened \cls may exceed this color.

Each \cl was observed on at least 2 different nights (except for NGC2682).
Each of these nights
was photometric (except in the case of NGC7078 and NGC6791, where the single
photometric night was used to calibrate the non-photometric night).
Thus, we obtained two independent measurements for each \cl (with the noted
exceptions).

Exposure times for the \st \cls
varied from a few seconds to a few minutes. Care was taken to avoid 
saturating the brightest \cl giants, which are needed to define the TRGB.
We typically 
obtained only a single exposure in each filter although occasionally several
exposures were obtained and the median was reduced. The airmass was 
almost always $<1.5$ and the seeing generally ranged from $1-2\arcsec$.
The \cl was usually centered on the CCD.

On each photometric night, a large number (typically 25-40) of \st stars from
the lists of Geisler (1990, 1996) were also observed, some more than once.
Care was taken to cover a wide color and airmass range for these standards in
order to properly calibrate the program stars.

\subsection{Reductions}

Each frame was trimmed, bias-subtracted and flat-fielded (using twilight
skyflats for C and either twilight skyflats or domeflats for \T1) using IRAF 
\footnote{IRAF is distributed by the National
Optical Astronomy Observatories, which is operated by the Association of 
Universities
for Research in Astronomy, Inc., under cooperative agreement with the National
Science Foundation} software.

The standard stars were measured with the aperture \phot routine APPHOT in IRAF.
Since all of the standards lie in \st fields which include $\sim 10$ \st stars
each, a mean value for the aperture correction (from an inner aperture $\approx$
FWHM to an outer aperture $\approx 7\arcsec$ in radius) was applied. 
Transformation equations of the form:

\begin{eqnarray}
c = C + a_1 + a_2\times (C-T_1) + a_3\times X_c
\end{eqnarray}
\begin{eqnarray}
t_1 = T_1 + b_1 + b_2\times (C-T_1) + b_3\times X_{T_1}
\end{eqnarray}

\noindent were used, where c and $t_1$ refer to instrumental magnitudes 
(corrected to 1s
integration using a zero point of 25.0 mag), C, $T_1$ and $(C-T_1)$ are standard
values, and
the appropriate air masses are given by X. We first solved for all three 
transformation
coefficients simultaneously (using the PHOTCAL package in IRAF)
for each night in a run
and derived mean color terms for each filter in that run.
We then substituted these
mean color terms into the above equations and solved for the remaining two 
coefficients
for each night simultaneously. 
The nightly rms
errors from the transformation to the standard system ranged from 0.017 to 
0.035 \mags in C and
0.010 to 0.020 \mags
in \T1, with means of 0.023 and 0.016 \mags, indicating the nights 
were all of good to 
excellent photometric quality.

The program 
\cl data were reduced with the stand-alone version of the DAOPHOT II 
profile-fitting program (Stetson 1987).
The standard FIND-PHOT-ALLSTAR procedure was generally performed 3 times on
each frame, with typically some 20000 objects being measured in each \cl.
A total of some 850,000 objects were photometered.
After deriving the \phot for all detected objects in each filter, 
a cut was made on the
basis of the profile diagnostics
returned by DAOPHOT. Only objects with $\chi <2$,
photometric error $<2\sigma$ more than the mean error at a given magnitude, and
$\mid Sharp \mid <0.5$ were kept in each filter (typically discarding about 
$10\%$ of the
objects), and then the remaining objects in the C and \T1
lists were matched with a tolerance of 1 pixel, yielding 
instrumental colors and \mags.
A quadratically varying PSF gave aperture corrections which were essentially 
constant with position except for data taken with the CTIO 0.9m 
before 1995  and data taken with the KPNO 4m before the new ADC was installed.
In both of these instances, quadratically varying aperture corrections were
required in addition to the quadratically varying psf.
Mean aperture corrections were determined
from  50-100 bright, unsaturated and
uncrowded stars (after subtraction of all other photometered stars).
The rms deviations about the mean
were generally only 0.014 \mag  in 
C and 0.010 \mag in \T1. These aperture corrections were then applied to all
remaining objects and \phot 
on the \st \Wash system was then obtained using the
above transformation equations. Since no photometric CCD data was obtained
for NGC2682, we used existing photoelectric data from Canterna (1976a) and
GCM for 10 stars in common to transform the CCD data to the \st system. 
These stars covered a wide color range that included the full range of the RGB.
Finally, the data of Friel and Geisler (1991) were used as
one of the two observations for NGC5927.

\subsection{Final Photometry and Errors}

We then generally had two independent observations on the \st system for each
\cl. In addition, most of the \cls also had photoelectric \phot available for
a small number of giants from previous studies.
The derivation of final colors and \mags was 
performed in the following manner. For the two
\cls (NGC6791 and NGC7078) for which only a single photometric observation was
available, these data were used to calibrate the non-photometric data, via 
a sample of several hundred
bright stars with a wide color range to determine the
zeropoint and color term in \T1 and \ct1.
The \phot was then averaged for all objects in all \cls
which had two observations (this
did not include all of the objects because of different field sizes and 
locations), leaving the \phot for single observations unchanged.     
For this procedure, we used Stetson's DAOMASTER routine, which provides the
mean difference $(\Delta)$ and standard deviation $(\sigma)$
for the objects in common. This mean
difference varied from 0.001 to 0.082 \mags for $\Delta(C)$, with a mean for
all of the \cls of 0.038 \mags, while the corresponding $\sigma$ values ranged
from 0.028 to 0.103, with a mean of 0.059 \mags. Likewise, the $\Delta(T_1)$
values ranged from 0.002 to 0.042, with a mean of 0.021
\mags, and the $\sigma(T_1)$
values ranged from 0.007 to 0.049, with a mean of 0.026 \mags. 

For the \cls which had existing photoelectric \phot (all except NGC5927), 
we then compared the CCD and photoelectric \phot for stars in common.
Although photoelectric \phot
of GC giants can suffer from crowding effects, the giants that had been 
observed were generally among the brightest in the \cl and were also selected
for relative isolation, minimizing such effects.
In addition, although crowding and photometric errors
may be more severe than for the CCD profile-fitting 
\phot, these \pe observations are not 
subject to systematic errors such as flat-fielding and aperture corrections
which can plague CCD data.
The results of this comparison are given in Table 2.
We list the mean difference (in the sense $CCD - PE), \sigma$ and the number
of objects in common for both \ct1 and \T1, as well as the source of the 
photoelectric \phot. Note that the comparison for NGC2682 is after transforming
the CCD data to the \pe system, so that the mean differences are 0 by 
definition. Also note that two of the \cls (NGC5272 and NGC6791) had no \T1
\pe \mags available. 

Figure 1 illustrates the resulting $\Delta(C-T_1)$ vs. $(C-T_1)_{PE}$ and
$\Delta(T_1)$ vs. $(T_1)_{PE}$ diagrams for a representative
\cl, NGC6397. In general, we find no significant
trends of $\Delta$ with either color or
\mag. This holds even for very red stars (\ct1$\sim 3.4$) in NGC6791, among the
reddest in our entire sample, which were observed using the $R_{KC}$ filter for
\T1, giving confidence that such observations are not subject to large 
systematic errors (although we do not have \pe data available for even redder
stars and thus cannot make such a statement for them). The mean differences
(excluding NGC2682 and NGC104, which has only 1 star) are: $<\Delta(C-T_1)>=
0.013\pm0.031(\sigma)$ and $<\Delta(T_1)>=0.000\pm0.022$. This is quite
satisfying agreement, indicating that our CCD \phot in general is very close
to the \pe system. However, in the case of NGC4590, NGC6397 and NGC5272, the
differences, especially for the colors, appear significant. For these 3 \cls,
we offset the CCD \phot by the indicated differences to be on the \pe system.

The final \Wash \T1 vs. \ct1 color-\mag diagrams (CMDs) for each \cl are shown
in Figure 2. The \phot can be obtained from the first author upon request.
From the above analysis, including nightly transformation errors, comparison
of different CCD frames and the comparison of the CCD and \pe data, we conclude
that this \phot is on the \st \Wash system with zeropoint errors of $\sim
0.03$ \mags. These \Wash CMDs show the same well-known features associated with
these \cls from studies in other photometric systems: the main sequence
and the turnoff, strong subgiant and RGBs,
the variation of HB
morphology with \met, the asympototic giant branch (AGB)
and a smattering of field
star contamination. 
The RGBs of the \cls
in particular are very well defined, except
for that of NGC5927, which suffers from severe field contamination  as well
as large and probably variable \red. However, a reasonable RGB is still 
visible. 
(Note that, given the small size of the CCD in comparison
to that of NGC2682, only a fraction of the \cl was covered and therefore we 
have supplemented the CCD data with the \pe data from Canterna (1976a) and
GCM.)
Of primary importance, it is clear that there is a very wide
range in the color of the RGB between a low \met \cl like NGC7078 and a high
\met \cl like NGC6791.

\section{The Standard Giant Branches}

Because of uncertainties in \reds and distances, it is best to derive the \sgbs
in the observational (\T1,\ct1) plane before transforming them to the absolute
\mag, dereddened color
(\MT1,\CT10) plane. Then, if the \red or distance scale changes,
they can be easily transformed again using the new values.

The analysis
was generally confined to the brightest $\sim 5$ \mags of the
RGB, which extends well below the level of the HB.
The \sgbs were manufactured by first excluding obvious HB, AGB and field stars.
Then a third order polynomial was fit to the remaining points, further
rejecting $>3\sigma$ outliers. The \cl RGBs are very populous: even after this
rejection, the fits involved an average of 361 stars in the 10 GCs. Note that
the DCA fits typically used $<50$ stars. For the less-populated open \cls,
only 14 giants were used in NGC2682 while 56 were available in NGC6791.
The \T1 \mag
was employed as the independent variable (which can slightly bias the fits
compared to a maximum likelihood solution). The rms of the
fits in \ct1 ranged from 
0.028 (NGC4590, NGC6397 and NGC7078) to 0.117 (NGC5927), with a mean rms of
0.058 \mags.
Note that this quantity does depend on the \met of the \cl, as the more 
\mp \cls have steeper RGBs. Although a third order polynomial was found to 
provide an excellent fit over virtually the entire RGB, some of the \cls (like
NGC6752, NGC104 and NGC5927) exhibit extreme curvature at the TRGB, in some
cases becoming fainter at the reddest colors, which was not well fit by the
software. In order to account for this additional curvature, a hand-drawn fit 
was made which extended from the upper well-fit part of the polynomial through
the reddest part of the RGB. In the case of NGC104, this hand-drawn curve was
terminated at \ct1$\sim 3.9$ although there are two stars lying much redder and
fainter that could well belong to the \cl GB.

The fits are shown in Figure 3 and the coefficients given in Table 3.
Only the fitted part of the RGB is shown. In
each diagram, stars included in the fit are represented by filled boxes,
excluded stars are open boxes, the large crosses
are \pe observations and the solid curve is the third order polynomial fit
(augmented by the hand-drawn curve for the TRGB in some instances). One can 
see again that the CCD data are generally in very good agreement with the \pe 
values. Note that the \pe
data were not used in the fit, except in the case of NGC2682.
The goodness of the fitted curves is also evident. 
The curves are of the form:
$(C-T_1)=a+b(T_1-O)+c(T_1-O)^2+d(T_1-O)^3$, where O is the \T1 offset also given
in the table.

To convert these fiducial GB curves to the (\MT1,\CT10) plane, a distance
scale and \cl \red must be adopted. We have utilized the Z85 \reds for the 
GCs; those for the open \cls are based on a mean of the best available values.
While the \reds for these \cls are all
relatively small and well known (except for that of NGC5927 and to a lesser
extent that of NGC6352), the question of the appropriate distance scale remains
controversial. This has become particularly true in the last year with the 
advent of the HIPPARCOS data. However, despite its intrinsic precision,
application of the HIPPARCOS data to derive GC distances has produced several
papers that are somewhat at odds (e.g. Reid 1997, Pont et al. 1998,
Chaboyer et al. 1998) and the last word on
the accuracy of this \tech is not yet in. For this reason and to maintain 
consistency with DCA, we have adopted their distance scale, which in turn is
based on the theoretical HB models of Lee et al. (1990). For these calculations,
the luminosity of GC RR Lyrae variables follows the relation: $M_V(RR)=0.82+
0.17[Fe/H]$. The variation of luminosity with abundance contained in this
relation, which is appropriate for a helium abundance of Y=0.23, agrees to
within the uncertainties with that derived from a variety of techniques, e.g.
Baade-Wesselink analysis of field RR Lyraes (Carney et al. 1992) and the
variation of HB \mag with \met for M31 GCs (Fusi Pecci et al. 1996). However,
as discussed in DCA, GCs with very red HBs and no RR Lyraes are problematical.
Theoretical models (e.g. Lee et al. 1990)
predict that the HB \mags of such \cls should be $\sim 0.1-
0.2$ \mags brighter than hypothetical RR Lyraes. Again, to maintain consistency
with DCA, we have adopted their value for this \mag difference of 0.15, i.e.
V(RR) is assumed to be 0.15 \mags fainter than the observed V(HB). Similar
values were obtained and used by  Ajhar et al. 
(1996) and Fusi Pecci et al. (1996). Thus, to obtain a \dm for 
GCs, we use the \feh
value from Z85 and the Lee et al. relation to derive $M_V(RR)$. We then subtract
this value from the observed V(HB) given in Armandroff (1989). For the
exclusively red HB \cls (NGC104, NGC6352 and NGC5927) we finally add 0.15.
For the open \cls, we simply used the mean of the best existing \dm 
determinations. Our \dm values are given in Table 1.

The \sgbs are presented in Tables 4 and 5.
Table 4 gives sample points at 0.1 \mag
intervals in \MT1 from \MT1$=+1$ to near
the TRGB, while Table 5 gives additional points near the TRGB.
Note that we have used $E(C-T_1)=
1.97E(B-V)$ and $A_{T_1}=2.62E(B-V)$ (Geisler et al. 1996). Also, for $A_V=3.2
E(B-V)$, we have $M_{T_1}=T_1+0.58E(B-V)-(m-M)_V$.

The \sgbs for all 12 \cls are displayed together in Figure 4. The \sgbs are
generally well separated, with similar shapes, and ranked in order of metal
abundance, with \met increasing from blue to red. It is also clear that the
\MT1 \mag of the TRGB is roughly constant for the most \mp \cls and then
increases with \met for the more \mr \cls. The color separation between the
\met extremes is very large. This is graphically portrayed in Figure 5, which
compares the \vi \sgbs from DCA
with the \ct1 \sgbs for the same \cls. The \Wash GBs
are much more widely separated than the \vi GBs.

\section{Metallicity Determination}

\subsection{Metallicity Calibration}

The main goal of this study is to develop a \tech which is a sensitive \met 
indicator and can be used to derive \met values in a wide range of applications.
Following DCA, we will calibrate the \CT10 color of the SGBs at a given \MT1
as a function of \met. The selection of the fiducial \MT1 value is of some 
importance. Clearly, as discussed by DCA and shown in Figure 4, the SGBs are 
more widely separated at brighter \mags. In addition, for application to distant
stellar systems, one would like to have this fiducial \mag as bright as 
possible, in order to obtain the most accurate \phot at this level. However,
the more \mr SGBs do not reach as bright an \MT1 as the more \mp \cls, and all
of the SGBs are less well defined along the upper RGB than at fainter \mags.
Therefore, the appropriate fiducial \mag may be a compromise. Note that DCA 
first selected \MI$=-3$, about 1 \mag below the TRGB, as their fiducial \mag.
However, in a subsequent paper (Armandroff et al. 1993), they developed the 
calibration for \MI$=-3.5$ due to the higher \met \sens,  brighter \mag and less
confusion with AGB stars. 

We have opted to derive \met calibrations for three different \MT1 values: 
--2.5, --2 and --1.5. The middle value represents a point roughly 1 \mag below
the TRGB for the \mp \cls and is therefore comparable to DCA's value of
\MI$=-3$. The brighter value will be more useful for distant, \mp systems while
the fainter value is generally better defined (more stars available). Indeed,
the most \mr GCs (NGC6352 and NGC5927) barely reach \MT1$=-2$ and the two open
\cls (NGC2682 and NGC6791) do not have stars at this \mag. We have extrapolated
the SGBs of these \cls to derive \CT10 at \MT1$=-2$ (and also slightly 
extrapolated that of NGC6791 to \MT1=$-1.5$). There may still be a significant
number of AGB stars present at this faintest \mag but by \MT1$\sim -2$ the 
AGB has generally blended with the RGB and should not have a significant effect
on the mean \ct1 color of the RGB.
It is clear from Figure 4
that the \Wash SGBs are still very well separated at even fainter \mags and that
useful calibrations could be derived even below --1.5. Indeed, unlike the case
of the \vi SGBs, the \ct1 SGBs still retain a very significant \met \sens even
below the HB (see Fig. 5).

The choice of \met scale is also important. While the Z85 scale for Galactic
GCs, which was used by DCA for their \met calibration, has been in 
vogue for many years and has generally held up well, several recent studies
suggest a different scale may be more appropriate. In particular, Carretta and 
Gratton (1997,  hereafter CG97) suggest, based on their high dispersion 
spectroscopic studies of a large number of GC giants, that the Z85 scale may be 
nonlinear with respect to the true \feh scale. The extensive study by
Rutledge et al. (1997) of Ca II triplet strengths supports the CG97 scale.

Again, we have opted to use three different \met scales: Z85, CG97 (as given
in Rutledge et al. 1997) and our own ``HDS" scale using the unweighted means
of all high dispersion spectroscopic studies listed in Table 3 of Rutledge et
al. but substituting the latest value of \feh(NGC7078)$=-2.40$ from Sneden et 
al. (1997) for their earlier value of --2.30 (Sneden et al. 1991), which was
based on many fewer stars. The \mets for the \st \cls on each of these scales
is given in Table 1. The Z85 \met for NGC1851 is --1.33 but a variety of studies
(e.g. DCA, Armandroff and Zinn 1988, Armandroff and Da Costa 1991, Geisler et
al. 1997) indicate that a more appropriate value is $\sim -1.15$ and that is the
value we have adopted.
Note that the latter two scales are genuine Fe abundance
scales, i.e. they measure Fe abundances directly, while the Z85 scale is a
``\met" scale involving many different techniques subject to different 
elemental abundances but generally Fe.
The temperature of the RGB
is mostly controlled by the primary electron donor   elements Mg and Si as well
as Fe. Any differences among the relative abundances of these elements in
different \cls could lead to significant temperature effects. This may in fact
be the reason for the differences between the Z85 and CG97 scales. 
Mg and Si are generally found to be enhanced  with respect to Fe in \mp GCs
and not enhanced in solar abundance stars. However, there are some \cls (e.g.
Pal 12 - Brown et al. 1997) that show different abundance patterns than normal
for their \met. Such details should be born in mind when deriving a \met using
the SGB and similar techniques.

Thus, we will derive 9 different calibrations: one for
each combination of fiducial \MT1 and \met scale, and the reader can choose
whichever they feel is most appropriate or take an average of the \mets derived
from different calibrations.
We also remind the reader that these calibrations depend on the \red and choice
of distance scale as well.

For each combination of \MT1 and \met, we derived both linear and quadratic
calibrations. 
The equations were of the form:
\feh$=a+b(C-T_1)_o$, and \feh$=a+b(C-T_1)_o+c(C-T_1)_0^2$.
The linear equation was adopted unless the rms of the quadratic
equation was significantly smaller. The coefficients, final number of \cls used
and the rms values of the
fits (in dex) are given in Table 6. All \cls available were used in all fits
with equal weight
except that NGC5927 was discarded from the (--2,Z85), (--1.5,Z85), (--2,HDS)
and (--1.5,HDS) fits because of its discordant position. This is not unexpected
given the wide RGB and substantial contamination and \red of this disk \cl.
The calibration curves are shown in Figure 6. The discarded NGC5927 points are
indicated in parentheses.

From the table and figure, it is clear that some of the calibrations are much
better defined than others. The lowest rms values at each \MT1 are obtained
for the Z85 scale, followed by the HDS scale. The CG97 calibrations for --2
and --1.5 are the poorest, with rms values of $\sim 0.12$ dex.
All three Z85 calibrations yield
equally small rms values, but the (--2,Z85) calibration is particularly 
impressive as it includes all of the \cls except for NGC5927 and still gives
an {\bf rms of only 0.04 dex} (even including NGC5927 the rms is only 0.07 dex).
This is our preferred calibration. Note that this is
the one most analogous to that of DCA for which they
obtained an rms of 0.07 dex using
8 \cls and a quadratic fit.
In the figure associated with this calibration we also plot the five
\cls from DCA (shown as plus signs) which are common to both samples.
Here we have added 1 to the $(V-I)_{0,-3}$ value from DCA
(to facilitate comparison with the \Wash values) and have used the Z85 \mets
for the DCA \cls, which in some cases are slightly different from the values
they used. The much higher \met \sens of the \Wash technique for the same
\cls is striking.

This is then a very powerful but also simple \tech for determining \mets.
After obtaining the \phot, the distance (on the Lee et al. 1990
scale) and \red (both of 
which can be determined with the same \phot - see Sections 5 and 6) are used
to place the data in the (\MT1,\CT10) plane. The \CT10 value at the fiducial
\MT1 value (as determined from a fit) is then used in conjunction with the 
appropriate calibration to derive a \met. In practice, some iteration may be
required, since the distance derivation may need an assumed \met.
Alternatively, the observed RGB can be compared directly to the \sgbs and a 
\met value interpolated.

\subsection{Comparison to Other Techniques}
\subsubsection{Comparison to the Washington Two-Color Diagram Technique}

The \Wash system has long been used to determine the \met of G and K giants.
The traditional use of this system, most recently described in GCM,
involves observations in 3 or 4 filters --
C,M,\T1 and $T_2$ -- where the \met is determined by the position of a giant
in a two-color diagram plotting a color index mostly sensitive to \met (e.g.
\ct1) vs. an index mostly sensitive to temperature (e.g. $T_1-T_2$). It is 
important to see how well the new \sgb \tech compares with the traditional 
\tech for determining \mets.

As discussed in GCM, one of the key measures of the suitability of a \tech for
determining \mets is its \met \sens, S, defined (Trefzger 1981) as the change
in the \met index for a given change in \met. For the two-color diagram 
\Wash \tech, S varies from a very low
0.04 (for the coolest, most \mp giants) to a very high 0.48 (for
warmer, solar-abundance giants) and also depends on which combination of filters
is used. For the new \sgb \tech, at least for the linear calibrations, S is
given by the inverse of the b coefficient in Table 6. For these linear 
calibrations, S is a constant, independent of \met or temperature. For our
preferred (--2,Z85) \cal, S=0.79. Thus, {\bf
the new \tech is some 1.6 to 20 times
more \met sensitive than the old \2cd \tech}. In addition, the coolest,
brightest giants fell outside the \met \cal of the \2cd \tech and could not be
used, despite their obvious advantage in terms of photometric precision. With
the new \tech, all stars along the RGB can be used.

Another important criterion of the utility of a \met determination \tech is
its \sens to \photic errors. For a given \photic error, the bigger its effect on
the \met determination the less useful the \tech. This criterion is also
discussed in GCM for the \2cd \tech. We use the same \photic error as they used,
namely $\sigma(C-T_1)=0.025$, which is a typical number, and $\sigma(T_1)=0.02$.
For the \2cd method,
typical \photic errors lead to \met errors of from 0.09 dex for intermediate
temperature, \mr giants to 0.83 dex for cool, \mp giants, again also depending
on the choice of color indices. For the \sgb \tech and our preferred \cal,
we derive a total \met error of only 0.034 dex. This is 2.7 to 25 times better
than the \2cd method, which is not unexpected given the relative \met 
sensitivities. If we use a more appropriate error of 0.03 in both indices,
the total \met error is still only 0.042 dex.

Thirdly, we also address the issue of how sensitive a \met index is to \red,
measured as the change in \met caused by a 
given change in the \red. GCM found that
an increase in the assumed \red of $\Delta E(B-V)=+0.03$ led to an increase in
the derived \met of from 0.02 to 0.60 dex, with the coolest, most \mp giants
again showing the greatest \sens. For this same \red increase, the \met derived
from the \sgb method is decreased by 0.12 dex (again using our preferred \cal.)
So in this case, for warm, solar abundance giants, the \2cd \tech is 
actually less \red
sensitive than the new \tech, while for \mp giants (\feh$<-1$) the new method
is much less affected by \red. 

These three comparisons show that, under most circumstances, 
{\bf the new \sgb
\tech is a much better \met indicator than the \2cd method}. For studies of approximately
solar \met stars, say in old open \cls where the \red is uncertain,
the \2cd is still competitive, with a smaller \red \sens but also smaller \met
\sens and more \sens to \photic errors. 
Also, the \2cd method does not require knowledge of the distance.
And, as discussed in 4.3, age effects using the new technique become 
important for objects younger than $\sim 5$ Gyr, which includes most open
clusters.
However, in all other instances, 
especially for \mets below \feh$\sim -1$, the new \tech is far superior in all
respects to the \2cd method. Clearly, the \sgb method also holds an important
edge in observational efficiency in view of the need to only observe in two
as opposed to three or four filters.

The study of very \mp objects will especially benefit from the use of the new
\tech. This is graphically illustrated by reexamining the \mets of 5 \mp GCs 
studied by Geisler et al. (1992). In this study, they used the \2cd method to
derive \mets of a number of \mp
Galactic GCs and found 5 of their sample -- NGC2298,
NGC4590, NGC4833, NGC5897, and NGC6101 -- to have 
surprisingly low derived \mets of $\sim -2.5$,
some 0.4 to 0.8 dex more \mp than their Z85 \mets, and more \mp than any other
GCs known. As emphasized by Geisler et al. (1992), the \reds of these \cls are
generally relatively poorly known and the \met errors large. 
They recognized the limitations of their
study and pointed to the need for further investigations, which subsequently
did indeed generally
confirm the Z85 \mets (McWilliam et al. 1992, Minniti et al. 1993,
Geisler et al. 1995). To illustrate the power of the new \tech, we have derived
\mets for these \cls using the \phot of Geisler et al. (1992) and the
preferred \cal of our \sgb \tech. (Note that 
NGC4590 is one of our \st \cls but that the \phot is different). We
find \mets of --1.71, --2.04, --1.97, --1.83, and --1.86 for NGC2298, NGC4590,
NGC4833, NGC5897 and NGC6101, respectively. These values compare very well with
their Z85 values, with a mean difference $<[Fe/H]_{SGB}-[Fe/H]_{Z85}>=-0.04\pm
0.11 (\sigma)$, compared to a mean difference of $-0.66\pm0.16$ using the \2cd
method. The spuriously low \mets derived from the \2cd \tech are undoubtedly due
to the lower \met \sens and higher \red and \photic error \sens of this method,
especially for low \met, cool giants.

\subsubsection{Comparison to the \vi Standard Giant Branch Technique}

We will also compare our \tech to the prototype  DCA \vi \sgb
\tech using the same 
criteria (and preferred calibration) as above. The \vi \tech has been in vogue
since DCA introduced it. Its popularity has even led to the selection of the
corresponding F555W and F814W filters on the WFPC2  on HST as
the standards for stellar \pop work with this important instrument.

The most direct comparison of the \met \sens of the two techniques is to 
compare the total difference in color index between the same GC 
\sgbs at a similar fiducial \mag. As noted above, there are 5 GCs in common to
the two studies, and \MI$=-3$ is roughly comparable to \MT1$=-2$. The total
difference in \VI0 between NGC104 and NGC7078 (the most \mr and \mp \cls in
DCA, respectively) at this \mag is 0.381, while these same \sgbs differ by
1.164 in \CT10. Thus, {\bf the \Wash \tech is more than 3 times as \met 
sensitive as the \vi \tech}.
Note that the actual \met \cal derived by DCA between these 
\cls is not linear, as assumed here, but quadratic. Thus, the \Wash \tech will
be even more sensitive at lower \mets but $<3$ times as sensitive at higher
\mets. If one instead uses a fiducial \mag that is 0.5 \mag brighter, the \Wash
\tech is only about 2.5 times as \met sensitive, still a substantial advantage.

To compare the relative \photic error sensitivities, we use the same \photic
errors as
above: 0.025 in \vi and 0.02 \mags 
in I. Since the effect of \mag errors is much 
less significant than that of color errors in this
regard, one expects that the relative
\photic error sensitivities will be similar to the \met sensitivities derived 
above. Indeed, one finds that the \Wash \tech is 2.9 times less sensitive to
\photic errors as the \vi
\tech at \MI$=-3$. In other words, {\bf for a given \photic accuracy, metal
abundances can be determined 3 times more precisely with \ct1 than \vi}.

To compare the \red sensitivities, we use $E(V-I)\sim 1.34E(B-V)$ and $A_I\sim
1.55E(B-V)$ for $R=A_V/E(B-V)=3.2$ (Dean et al. 1978). For an increase in the
assumed \red of 
$\Delta E(B-V)=+0.03$, the derived \vi \met is decreased by about 0.19 dex,
compared to 0.12 dex for the \ct1 \met. So {\bf
the \Wash \tech has only about half
of the \red \sens}, in terms of its effect on \met, as the \vi method.

Thus, the \Wash \tech enjoys a significant advantage over the corresponding \vi
\tech for determining accurate \mets. However, it is important to also compare
their observing efficiencies. We determined the \met accuracies achieved in a
given time for a given \mag (namely 1 \mag below the TRGB)
with the two techniques. We investigated both a \mp
(\feh$=-2$) giant and an intermediate \met
(\feh$=-1$) giant, and used the appropriate
colors based on the \sgbs. We employed the count rates
given in the latest KPNO Direct CCD Observing Manual for the KPNO 4m telescope,
new ADC corrector and T2KB CCD. We assumed no moon and $1\arcsec$ seeing in
all filters except for $1.2\arcsec$ in C. In a total integration time of 1 hour
on an intermediate \met giant with $V=23.6$, a \photic error corresponding to
a \met error of 0.10 dex was achieved in \vi while a \met error of only 0.36
dex was obtained in \ct1. However, for the \mp giant, the roles were reversed,
with \vi achieving only a 0.23 dex \met error while \ct1 yielded an error of
0.20 dex. Of course, this does not take into account aperture correction errors
(which would tend to favor \ct1), errors in transformation to the \st system
(which would tend to favor \vi), etc. But in general it appears that \vi is
more efficient at obtaining precise \mets for objects with \feh \gtsim --1.5
while \ct1 wins for more \mp objects. Note that further improvements in the
quantum efficiency of CCDs in the blue-uv will help to improve the C \sens,
but similarly care must be taken to maintain maximum throughput in the optics
in this spectral region. 

We note that \sgbs do exist in $V,(B-V)$ (Sarajedini and Layden 1997). 
They find that the NGC104 and NGC7078 \sgbs are separated by 0.743 \mags in
$(B-V)_0$ at $M_V=-2$, so the corresponding  $V,(B-V)$ \met \sens is slightly
$<2/3$ that of the \Wash system.
A comparison with their \met calibrations used in deriving the
equivalent of the SRM method shows that, for their 6 calibrating \cls, they
obtained rms values of from 0.05 to 0.08 dex, roughly comparable to our values.

\subsection{Age Effects}

The well known degeneracy between age and \met effects for stellar \pops must
be addressed. The \st \cls comprising our sample range in age from $\sim 4$ Gyr
for NGC2682 to $\sim 10-15$ Gyrs for the other \cls (depending on the distance
scale adopted for the GCs and the selected isochrones).
The two old open \cls were added to establish the
\mr end of the \met \cal, with the hope that age effects would be small. How
well was this hope born out? Clearly, given the very old age ($\sim 10$ Gyr)
of NGC6791, age effects relative to those for the GCs should be minimal. And
indeed a glance at Figure 6 verifies this. What about the much younger \cl
NGC2682? Figure 6 shows that M67 does indeed lie at a slightly bluer color than
expected for its \met in all calibrations in which it is involved. However, the
effect is small, amounting to only about 0.1 dex or less. Such an effect could
actually be due to our assumed \met being slightly too high. But it is 
consistent with the younger age causing a bluer RGB and leading to an 
underestimate of the \met. 

Thus, empirically
it appears that our \met calibrations may be safely applied to any \cl
older than $\sim 4$ Gyr with 
only a small effect on the derived \met. An analysis
based on Washington  isochrones (Lejeune 1997) supports this.
These isochrones include $UBVRIJHK$ as well as $CMT_1T_2$ and thus results
from different color systems can be directly compared.
An indication of the reliability of these isochrones is given by the fact that
they yield  a color difference in \CT10 (at \MT1$=-2$) that is 3.16 times larger
than the corresponding \VI0 difference (at \MI$=-3$) between \feh=$-2.2$ and 
$-0.6$ (15 Gyr) isochrones, in excellent accord with the results found
in the last section for the differences between the NGC104 and NGC7078 \sgbs.
The 8 and 15 Gyr, \feh$=-1.6$ isochrones are separated by 0.093 \mags in
\CT10 at \MT1$=-2$ and by 0.028 \mags in \VI0 at \MI$=-3$. These color
differences would lead to very 
similar \met ``errors" of 0.12 and 0.11 dex, respectively, if gone unrecognized.
Similarly, the differences between 8 and 15 Gyr, \feh$=-0.6$ isochrones are
0.231 and 0.071, leading again to very similar \met errors of 0.29 and 0.27
dex, respectively. 
As quoted in DCA, the Revised Yale Isochrones (Green et al. 
1987) indicate that a 7 Gyr, \feh$=-1.3$ isochrone is 0.05 \mag bluer in
\VI0 at \MI$=-3$ than a 15 Gyr isochrone of the same \met, leading to a \met
error of 0.2 dex, in very good agreement with the trend from the above 
isochrones.
So the age
\sens of the \Wash \sgb \tech
appears to be very similar to, and perhaps slightly higher than,
that of the \vi method for old \cls, and both systems are more affected by
age at higher than lower \mets.

What about for even younger \cls? Clearly, by a certain age, the RGB will be
moved sufficiently to the blue relative to an older \cl of the same \met that
the effect on the derived \met will be significant.
The Lejeune isochrones indicate that 
the age \sens becomes significant for \cls 5 Gyr and younger, especially for
the \Wash system.
A similar value is obtained from the recent analysis
of Bica et al. (1998) where they compared \mets derived from the \Wash \sgb
\tech with those available from spectroscopic studies for 5 Galactic open \cls
and 6 LMC \cls whose ages ranged from 1--4 Gyrs (most were $\sim 2$ Gyr).
For this sample, a clear trend
was found for the \Wash \sgb \mets to underestimate the spectroscopic \met by
an approximately constant amount, independent of age.
An unweighted mean yielded a difference of
$0.41\pm 0.21$ dex. Thus, the \sgb \tech derived here should only be applied
to \cls older than $\sim 5$ Gyr.

We note in passing here that the Lejeune isochrones can also be used to test
how well the \Wash system works for deriving ages from main sequence \phot.
VandenBerg et al. (1990) and Sarajedini and Demarque (1990) have shown that
the color difference between the turnoff and the lower subgiant branch is a
sensitive and powerful age indicator. For a \feh$=-0.6$ isochrone, this color
difference is $\sim$0.27 and 0.17 \mags in \VI0 for a 10 and 20 Gyr population,
while in \CT10 the respective values are 0.71 and 0.47 \mags. Thus, the 
difference in these two values is some 2.5 times larger in \CT10 than \VI0,
indicating that one could indeed obtain significantly
greater age accuracy using the \Wash
system for a given \photic accuracy. These same age isochrones are separated
by $\sim$0.13 \mags in both color indices for \feh=$-2.2$ so for these lower
\mets the age \sens is the same. The \Wash system thus holds great potential
for deriving accurate ages as well as \mets.

\subsection{Intracluster Metallicity Dispersion}

As developed by DCA, one can use the dispersions in the fit of the \sgbs to
the data to derive upper limits to the intrinsic \met \disp in each \cl.
In order to investigate this quantity, we have used all \cl stars which fell
between \MT1$=-1.5$ and --2.5, only excluding those stars that fell far away
from the RGB. The standard devation, $\sigma(C-T_1)$, about the fitted \sgb was
calculated and this value was converted into a \met \disp using our preferred
\cal. The results are displayed in Table 7, where we give the number of stars
and the standard deviations in color and \met. Note that the two open \cls did
not have sufficient stars in this part of the RGB to be useful. Also note that
this is an upper limit to the intrinsic \met \disp, as we have not taken 
\photic errors or contributions by AGB stars or field stars into account. 

The results generally show very small upper limits to the intrinsic \met \disp
of our \st \cls, with a typical limit of $\sim 0.06$ dex. The limit for NGC5927
is especially large due to the presence of significant field star contamination
as well as differential \red. Our limits are generally similar to or lower than
those derived for the same \cls by DCA, who  found limits of 0.04 dex 
for NGC104, 0.07 for NGC1851, 0.06 for NGC6397, 0.05 for NGC6752, and 0.09 for
NGC7078. Recall that our observations were generally centered on these crowded
\cls, whereas the DCA data were generally offset, leading to increased \photic
error in the former with respect to the latter.

Such a procedure can be used to determine the \met \disp in program objects.
In this instance, the measured \photic error can be subtracted in quadrature 
from the observed scatter to derive a realistic estimate for any intrinsic 
\disp. Indeed, given a sufficient sample in a system such as a dwarf
spheroidal galaxy in which an intrinsic \met \disp is expected,
one could even derive the \met distribution of the giants using this \tech, as
was done by Geisler and Sarajedini (1996).

\section{Simultaneous Reddening and Metallicity Determination}

As noted in the Introduction, Sarajedini (1994) devised the simultaneous
reddening and metallicity (SRM) technique to facilitate the determination
of these quantities in an internally consistent manner. The SRM method
exploits the fact that the shape and position of the RGB are dependent
on metallicity and reddening. In addition to the calibration of these
quantities using standard RGB sequences, the SRM method also requires
knowledge of the HB magnitude, the color of the RGB at the level of
the HB, and the shape of the RGB.

The first step in establishing the SRM calibration is to estimate the
value of $T_1(HB)$ for each cluster. This is a rather complicated endeavor
because of the diversity of HB types among the clusters. For the clusters
with RR Lyraes, we proceed as follows. Working in the ($T_1,C-T_1$) plane,
we fit a cubic polynomial to the HB stars that straddle the RR Lyrae
instability strip. This is done using an iterative 2$\sigma$ rejection
algorithm similar to that utilized for the RGB fits above. Based on the
work of Sandage (1990), we have that the color at the blue edge of the
RR Lyrae instability strip is $(B-V)_0 = 0.18$ and the color of the
red edge is  $(B-V)_0 = 0.40$. Converting these colors to $(C-T_1)_0$
with the transformations of Geisler (1996) gives a mean color of
$(C-T_1)_0 = 0.45$ for the center of the instability strip. For those
clusters with RR Lyraes, we read off $T_1(HB)$ from the
polynomial fits at $(C-T_1)_0 = 0.45$. The rms of the fit added in
quadrature with the uncertainty in the photometric zeropoint of
0.03 mag is adopted as the error in $T_1(HB)$.

In the case of the two clusters with purely blue HB morphologies, NGC6397 and
NGC6752, we use a different approach. First, we fit a polynomial to
the blue HB stars in NGC5272. We then shift the HBs of NGC6397 and NGC6752
in $T_1$ until the mean star-by-star difference with the NGC5272 blue HB fit is
minimized. The color shift is set by the difference in reddening between
each cluster and NGC5272. Then, knowing $T_1(HB)$ for NGC5272 from the
procedure described above and the shift required to match the blue HBs,
we can infer the values of $T_1(HB)$ for NGC6397 and NGC6752. The error
is estimated by adding in quadrature the uncertainty in the $T_1(HB)$ value
of NGC5272 and the uncertainty in the shift.

There are 4 clusters with purely red HBs in our sample. We are omitting
NGC2682 from the SRM calibration because its RGB is too sparsely populated.
In the case of these clusters, we construct $T_1$ mag histogram distributions
of the HB stars. Fitting a gaussian curve {\it in the region of the peak} in
these histograms yields the value of $T_1(HB)$, while the error in
$T_1(HB)$ is given by the dispersion in the fitted gaussian divided by
the square root of the number of points used in the curve fit. The resulting
error ends up being quite small (typically $<0.01$ mag) because of the
large number of points on the red HB. The only significant error in the
$T_1(HB)$ of the red HB clusters is that associated with the photometric
zeropoint, which we estimate to be $\sim$0.03 mag. We note in passing that
fitting the Gaussian in the region of the histogram peak minimizes 
the uncertainties in the value of $T_1(HB)$ introduced by the evolution of 
stars away (brightward) from the Zero Age Horizontal Branch.

Once we have settled on values of $T_1(HB)$ for each cluster, then we
utilize the polynomial RGB fits to read off the $(C-T_1)_g$ value, which
is converted to $(C-T_1)_{0,g}$ by applying our adopted reddenings. We can
also construct the magnitude difference between the HB and RGB at
$(C-T_1)_0 = 2.4$ (i.e. $\Delta$$T_{2.4}$). Both of these quantities,
$(C-T_1)_g$ and $\Delta$$T_{2.4}$ vary with metal abundance, and are
listed in Table 8 along with the other measured parameters. Performing
a weighted least squares fit gives us the two relations that are central to
the SRM method,

\begin{eqnarray}
[Fe/H] = a_1 + b_1\times(C-T_1)_{0,g} + c_1\times(C-T_1)_{0,g}^2 + d_1\times(C-T_1)_{0,g}
^3
\end{eqnarray}
\begin{eqnarray}
[Fe/H] = a_2 + b_2\times \Delta T_{2.4} + c_2\times \Delta T_{2.4}^2 + d_2\times \Delta T
_{2.4}^3.
\end{eqnarray}

Table 9 gives the values of these coefficients for the various metallicity
scales considered herein, while Fig. 7 illustrates the relations. The 
metallicity errors are taken from Zinn \& West (1984) for the Z85 abundances
and Rutledge et al. (1997) for the CG97 and HDS abundances. 

The reader is referred to the work of Sarajedini (1994) for a detailed
description of how to apply the SRM method. In summary, one needs a
polynomial or fiducial representation of the cluster RGB sequence as well
as estimates for the observed values of $T_1(HB)$ and $(C-T_1)_g$.
Then, an iterative procedure can be set up that utilizes these observed
properties in conjunction with the two equations
above to provide estimates of the cluster reddening and metallicity.
In addition, Monte Carlo simulations are used to determine the uncertainty
in these quantities. Given a well-determined RGB sequence with errors of
0.03 mag in $T_1(HB)$ and 0.03 mag in $(C-T_1)_g$, we expect metallicity
errors of 0.15 dex in $[Fe/H]$ and reddening errors of 0.05 mag in 
$E(C-T_1)$ (0.025 mag in E(B-V)). As for the effects of cluster age, Sarajedini
\& Layden (1997) and Mighell et al. (1998) have shown that the SRM method
is insensitive to age for clusters $\sim$4 Gyr and older.

\section{Additional Metallicity, Reddening and Distance Determination 
Techniques}

\subsection{Metallicity Determination from the 
Slope of the Red Giant Branch}

The work of Hartwick (1968) illustrated the suitability of the RGB slope
as a metallicity indicator. In the present work, we define the RGB slope
($S_{-2}$) as
\begin{eqnarray}
S_{-2} = \frac{-2.0}{[ (C-T_1)_g - (C-T_1)_{-2} ]},
\end{eqnarray}
where $(C-T_1)_{-2}$ is the color of the RGB at 2 magnitudes above the HB.
This quantity is well correlated with metal abundance; as such, we can
construct the following relation -
\begin{eqnarray}
[Fe/H] = a + b\times S_{-2} + c\times S_{-2}^2
\end{eqnarray}
for the three metallicity scales. Table 10 lists the values of these
coefficients while Fig. 8 illustrates the relations. The RGB slope
method of metallicity determination is especially useful since it does not
require knowledge of the reddening or the distance. Furthermore,
Mighell et al. (1998) have shown that, in the $B-V$ passbands, it
is insensitive to age for clusters older than $\sim$4 Gyr. If the
values of $(C-T_1)_g$ and $(C-T_1)_{-2}$ can be determined with an
error of $\pm$0.03 mag, then the resultant error in $[Fe/H]$ is
approximately 0.08 dex.

\subsection{Metallicity Determination from the Magnitude of
the Red Giant Branch Bump}

As a star evolves up the first ascent RGB, it reaches a point where
its evolution pauses or reverses course for a short time, after which
it resumes its brightward movement in the H-R Diagram
(Thomas 1967; Iben 1968). This phenomenon
leads to a clumping of stars along the RGB. When a luminosity function
(LF) is constructed, this clump manifests itself as a bump in the LF,
hence the name `RGB bump.'

From theoretical considerations, the luminosity of the RGB bump is
dependent on age and abundance (Iben 1968, Fusi Pecci et al. 1990). However,
since the absolute zeropoint of the theoretical RGB bump luminosity
is uncertain, it is more useful to measure the RGB bump
magnitude relative to that of the HB (i.e. $\Delta$$T_1(Bump-HB)$).
In these terms, as a cluster's age and/or metallicity increases,
$\Delta$$T_1(Bump-HB)$ also increases.
For clusters that are older than $\sim$10 Gyr, the effect of age on
the absolute magnitude is negligible (Alves \& Sarajedini 1998); as a
result, we can
utilize the magnitude difference between the RGB bump and the HB
as a metallicity indicator as proposed by Fusi Pecci et al. (1990)
and parameterized by Sarajedini \& Forrester (1995).

To isolate the location of the RGB bump, we proceed in the same manner
as Sarajedini \& Norris (1994). First, we construct a cumulative LF of
the RGB stars. As illustrated in their Fig. 15, the RGB bump clearly
stands out in such LFs. After fitting a `continuum' to the LF around
the region of the bump, we subtract this off resulting in a flattened
RGB LF. The maximum point in this flattened LF represents the faintest
extent of the RGB bump, while the zero crossing immediately brightward
represents the onset of the bump as one proceeds fainter from the RGB
tip. We therefore adopt the magnitude midway between these two points as the
value of $T_1(Bump)$, which is then coupled with $T_1(HB)$ to produce
$\Delta$$T_1(Bump-HB)$. The error in $T_1(Bump)$ is half the distance
between the maximum and the zero-crossing multiplied by 0.68 to simulate
a 1-$\sigma$ uncertainty.

As in the case of the RGB slope, we parameterize the variation of the
RGB bump with metallicity using the following relations.
\begin{eqnarray}
[Fe/H] = a + b\times \Delta T_1 + c\times \Delta T_1^2,
\end{eqnarray}
where $\Delta$$T_1$ is the difference in magnitude between the bump and
the HB; the fitted coefficients are listed in Table 11. Figure 9 shows
the fitted points and the resulting relations.
If the value of $\Delta T_1$ can be measured to  $\pm0.1$ \mag, then the
\met determined  from the RGB Bump
will have an uncertainty of $\sim$0.1 dex.

Our CMDs are sufficiently populous that, at least in several instances,
we see evidence of a similar LF enhancement on the AGB $\sim 1$ \mag above
the HB. The best cases are NGC104, NGC5272 and NGC7078. We believe that this
feature is the AGB bump, due to a phenomenon similar to that which produces
the RGB bump. Such a feature has recently been identified in populous LMC
field star CMDs (Gallart 1998) and may well be responsible for the `VRC'
suggested by Zaritsky \& Lin (1997) as being due to a possible foreground 
galaxy.

\subsection{Reddening Determination from the Red Edge of the Blue Horizontal
Branch}

The temperature limits of the RR Lyrae instability strip are well-defined,
both theoretically and observationally (see Sandage 1990). Previous
studies have exploited this fact to derive the \red of a \cl, given a large
sample of RR Lyraes and giants along the BHB: the observed color limit between
the variables and non-variables is compared to the intrinsic color limit,
directly yielding the \red. We have undertaken a similar analysis, using the
six \st \cls (NGC1851, NGC4590, NGC5272, NGC6362, NGC6752 and NGC7078) which
have both RR Lyraes and BHB stars. NGC6752 in fact does not have RR Lyraes 
but does appear to have BHB stars that lie very near the edge of the instability
strip. On the other hand, the reddest
BHB stars in NGC6397 fall substantially blueward
of the instability strip.

It is clear from Figure 2 that the BHB stars in these \cls
generally lie in a fairly     
tight sequence, that the RR Lyraes fall in a more scattered distribution from
fainter, redder colors to brighter, bluer colors, and that the 
color limit between 
these two types of stars is reasonably well defined. We derived this limit in
\ct1 and converted it to \CT10 using the \cl \reds. The six \cls have 
$<(C-T_1)_o>=0.17$, with a $\sigma$ of only 0.03. Sandage (1990) gives this
limit as $(B-V)_0=0.18$ which is \CT10=0.22 using the conversion of Geisler
(1996). Giving some weight to this determination, 
we adopt $(C-T_1)_0=0.18\pm0.04$ for the intrinsic color of the red
edge of the BHB. We can then use this value to derive the \red to a program
object which has a sufficient number of BHB and RR Lyrae stars, with an 
estimated error of $\sigma(E(B-V))=0.025$. This \tech can be used to 
supplement the \red derived from the SRM method described above.

\subsection{Distance Determination from the \T1 Magnitude of the
Tip of the Red Giant Branch}

The I \mag of the TRGB has become an increasingly popular standard candle in
recent years. The work of Lee et al. (1993) and others has shown
that this is indeed a very useful distance indicator. We have investigated the
analogous use of the \T1 \mag of the TRGB as a distance indicator.

A glance at Figure 4 suggests that $M_{T_{1_{TRGB}}}$ has at most a very small 
dependence on \met for \mp \cls (\feh \ltsim --1.15). We have determined this
value for the 6 \st \cls falling in this regime. A mean  $M_{T_{1_{TRGB}}}
=-3.22\pm
0.11(\sigma)$ is obtained. The small rms indicates that this should indeed be
a useful distance indicator for such objects. Clearly, though, at higher \mets
$M_{T_{1_{TRGB}}}$ 
increases rapidly with increasing \met and is not useful. Indeed,
such behavior is also seen for $M_{I_{TRGB}}$ for \feh \gtsim --0.75.

We can use the Bertelli et al. (1994) isochrones to investigate theoretical
predictions concerning how  $M_{T_{1_{TRGB}}}$ depends on age and \met.
Geisler (1996) has shown that the \T1 and $R_{KC}$ \mag scales are virtually
identical, with an almost negligible zeropoint (0.003) and color term (0.017)
relating them. Therefore, we can simply use the $M_{R_{KC}}$ \mags generated by
Bertelli et al. and compare them directly to our \MT1 results. The results are
shown in Figure 10 where  $M_{T_{1_{TRGB}}}$ 
is plotted vs. \feh. The plus signs
are from the Bertelli 
et al. models, where we have used an age of 12 Gyrs except for
the point at solar \met (4 Gyr) and \feh$=+0.4$ (10 Gyr) to overlap with the 
ages of our \st open \cls at these \mets. The squares are from our \sgbs. The
agreement between the models and observations is excellent. Both sets of data
indicate that   $M_{T_{1_{TRGB}}}$
is very sensitive to \met for \feh \gtsim --1.2
but that for more \mp \cls there is only a slight or possibly negligible \met
dependence. Therefore, for such \mp objects,  $M_{T_{1_{TRGB}}}$ can indeed be 
used as an accurate distance indicator, with a value of --3.22
and an error of \ltsim 0.15 \mags, for ages from $\sim 3-20$ Gyr.

\section{Summary}

We have obtained CCD
\phot in the \Wash system for a very large sample of stars in
each of 10 Galactic GCs and 2 open \cls. The \phot is on the \st system to
within \ltsim 0.03 \mags.
We have fit third-order
polynomials to each \cl to derive the \sgb. These \sgbs are converted to the 
(absolute \mag, dereddened color) plane by assuming the Lee et al. (1990) 
distance scale and \reds from Z85. We then derive \met calibrations
for the \CT10
color at three different fiducial \MT1 values, and for three \met scales. Our
preferred \cal is for \MT1=--2, about a \mag fainter than the TRGB
of metal-poor clusters, using the
Z85 \met scale. This \cal is very analogous to that derived by DCA for VI \phot.
We find that the \Wash system enjoys three times the \met \sens of the VI \tech.
It is also less sensitive to \red. The \Wash \sgb \met \tech is also superior
to the \Wash \2cd \tech in virtually all respects. The \sgb \tech is immune to
age effects for objects older than $\sim5$
Gyr. We derive upper limits of typically
0.06 dex for any intrinsic \met dispersion in the \st \cls.

We also use the \sgbs to derive a method analogous to that of Sarajedini (1994)
for determining both the \met and \red simultaneously. The \mag difference
between the HB and the RGB bump, and the slope of the RGB,
are also found to be sensitive \met indicators. 
In addition, \red can be determined from the color of the red edge of the 
blue HB. Finally, the \T1 \mag of the TRGB is an accurate distance indicator
for objects more \mp than \feh$\sim -1.2$ and older than 3 Gyr.
An analysis of available isochrones indicates that
the \Wash system also holds great potential
for deriving accurate ages as well as \mets.

Gary Da Costa  and Taft Armandroff provided much of the inspiration for this work
from their original study in the V,I system (DCA).
We thank Hugh Harris for a critical reading of a preliminary version of this manuscript.
The authors would like to thank Juan Clari\'a and
Cristina Torres for providing important photoelectric data 
in advance of publication. Juan Clari\'a and Andr\'es E. Piatti also very kindly
obtained one of the CCD fields for us. Pat Durrell allowed us to obtain 
another field during telescope time set aside for his thesis.
Peter Stetson, as always, has been very generous in allowing the use of his own
very useful programs. Jim Roberts and Eva Grebel are acknowledged for their 
initial work on another project which helped to launch the present one.
The \cl \phot can be obtained from the first author upon request.
This research is
supported in part by NASA through grant No. GO-07265.01-96A (to DG) 
and HF-01077.01-94A (to AS) from the 
Space Telescope Science
Institute, which is operated by the Association of Universities for
Research in Astronomy, Inc., under NASA contract NAS5-26555.
This research is supported in full by E. Geisler through her constant love and
encouragement.

\newpage

\centerline{\bf Figure Captions}
\begin{figure}
\figcaption{a. The difference between the \ct1 value derived from CCD and 
photoelectric \phot, as a function of the photoelectric value, for giants in
NGC6397 in common. b. As for a). except for the \T1 \mag.}
\end{figure}

\begin{figure}
\figcaption{\T1 vs. \ct1 color-magnitude diagram for the \st \cls. a. NGC104
b. NGC1851 c. NGC2682 NB - in this diagram, CCD observations are denoted by
squares and photoelectric data by 
crosses. d. NGC4590 e. NGC5272 f. NGC5927
g. NGC6352 h. NGC6362 i. NGC6397 j. NGC6752 k. NGC6791 l. NGC7078}
\end{figure}

\begin{figure}
\figcaption{\T1 vs. \ct1 CMD for the \st \cls showing the \sgbs. 
Filled squares 
are observations used in the fit, open squares are unused or
rejected stars. Large crosses indicate photoelectric observations. The solid
curve is the \sgb fit. a. NGC104 b. NGC1851 c. NGC2682  d. NGC4590 e. NGC5272 
f. NGC5927 g. NGC6352 h. NGC6362 i. NGC6397 j. NGC6752 k. NGC6791 l. NGC7078}
\end{figure}

\begin{figure}
\figcaption{The \Wash \sgbs in the (\MT1,\CT10) plane. At \MT1=--1.5, the \sgbs
are (from left to right): NGC7078, NGC4590, NGC6397, NGC5272, NGC6752, NGC1851,
NGC6362, NGC104, NGC5927, NGC6352, NGC2682 and NGC6791.}
\end{figure}

\begin{figure}
\figcaption{A comparison of the same VI and \Wash \sgbs. The VI data are taken
from DCA and plot \MI vs. $(V-I)_0$. The \Wash data plots \MT1 vs. \CT10.
The \cls are (from left to right): 
NGC7078, NGC6397, NGC6752, NGC1851 and NGC104. The \Wash \sgbs are much more 
widely separated than the VI RGBs.}
\end{figure}

\begin{figure}
\figcaption{Metallicity calibrations for various combinations of fiducial \mag
and \met scale. a. (\MT1=--2.5,Z85) b. (--2.5,CG97) c. (--2.5,HDS) d. (--2,Z85).
This is our preferred \cal.
The discarded point (NGC5927) is shown in parantheses. The plus signs show the
DCA \sgbs, using \MI=--3 and adding 1 to \VI0.
e. (--2,CG97) f. (--2,HDS) g. (--1.5,Z85) h. 
(--1.5,CG97) i. (--1.5,HDS)}
\end{figure}

\begin{figure}
\figcaption{Calibration of the SRM method for deriving \red and \met 
simultaneously. \feh (for three different \met scales) is shown as a function 
of the CMD parameters $(C-T_1)_{0,g}$ and $\Delta T_{2.4}$ for the \st \cls.}
\end{figure}

\begin{figure}
\figcaption{Calibration of the RGB slope  for deriving \met.
\feh (for three different \met scales) is shown as a function 
of the RGB slope.}
\end{figure}

\begin{figure}
\figcaption{Calibration of the RGB bump  for deriving \met.
\feh (for three different \met scales) is shown as a function 
of $\Delta T_1$(Bump-HB).}
\end{figure}

\begin{figure}
\figcaption{\MT1 at the RGB tip as a function of \met. The squares are the 
\sgbs; the plus signs are from the model isochrones of Bertelli et al. (1994).
For \mets \ltsim -1.2, $M_{T_{1_{TRGB}}}$ is nearly constant and can be used 
as a distance indicator.}
\end{figure}

\end{document}